# Investigation of the influence of temperature on the conductive properties of copolymer PVC -PolyAcetylene films

D.V. Vlasov , V.I. Kryshtob , T.V.Vlasova, L.A. Apresyan, S.I. Rasmagin

Prokhorov General Physics Institute, Russian Academy of Science

119991, Moscow, Russia

## ABSTRACT

The temperature dependence of conductivity of partially dehydrochlorinated PVC films, containing in their macromolecules chains of polyene-conjugated bonds (PCB) and representing copolymer PVC-Polyacetylene. In samples with excess of some "threshold" concentration of PCB with increasing temperature it was found conductivity switching on 10 -11 orders of magnitude. Instability of states with high conductivity in the temperature range which depends on the concentration of PCB was detected. Qualitatively, the increase of concentration of PCB was monitored by fixing the fluorescent and absorption spectra.

## Исследование влияния температуры на электропроводящие свойства образцов пленок из сополимера ПВХ-ПАц.

Д.В.Власов, В.И.Крыштоб, Т.В.Власова, Л.А.Апресян, С.И.Расмагин

### Аннотация

Исследована температурная зависимость проводимости образцов пленок из частично дегидрохлированого ПВХ, содержащего в цепи своих макромолекул полиен-сопряженные связи (ПСС) и представляющего собой сополимеры ПВХ-Полиацетилен (ПАц) В образцах с превышением «пороговой» концентрации двойных сопряженных связей обнаружены переключения проводимости на 10-11 порядков величины по мере увеличения температуры. Обнаружена неустойчивость состояния с высокой проводимостью в интервале температур, ширина которого зависит от концентрации ПСС. Качественно увеличение концентрации ПСС контролировалось посредством фиксации люминесцентных спектров и спектров поглощения.

### 1.Введение

Температурная зависимость проводимости является одним из наиболее информативных методов выявления физических механизмов переноса зарядов и проводимости, в особенности, когда исследуются полимерные и композитные материалы. В частности, как ионный механизм Френкеля – Шоттки [1,2] , так и прыжковые механизмы проводимости в модели Мотта удобно описывать феноменологической формулой для удельного сопротивления $\rho$

$$\rho = \rho_0 \exp\{-(\frac{W}{T})^\gamma)\, , \qquad (1),$$

где константы $\rho_0$ и W определяются из эксперимента, а коэффициент γ определяет наклон прямой в координатах $\ln(\rho\sigma)$ <=> 1000/T. При этом, если прыжковый механизм Мотта [2] реализуется в исследуемом образце, то γ =1/4, тогда как для механизма Френкеля Шоттки — γ =1, как, впрочем, и для обычного больцмановского распределения энергий зарядов. Известны и другие модели переноса заряда с другими значениями коэффициента [2].

В металло-подобных материалах с частично заполненной зоной проводимости изменение температуры обнаруживает положительный Температурный Коэффициент Сопротивления (ТКС). В полупроводниках ТКС, как правило, отрицателен, поскольку увеличение температуры приводит к появлению частиц носителей заряда с большими энергиями, которые уже могут переходить в зону проводимости, что и приводит к радикальному увеличению подвижности зарядов и соответственно увеличению частоты и дальности прыжков.

Для широкозонных полимерных материалов со свойствами, близкими к изоляторам, традиционно предлагается ионный тип проводимости, причем повышение температуры приводит к высвобождению зарядов (в том числе и ионов) из дефектных узлов, ловушек и примесей. Характерная зависимость сопротивления термопластичного полимера [1] может в координатах $LOG(\sigma)$ <=>1000/T иметь кусочно-линейный вид, причем участки, соответствующие стеклообразному и вязко-текучему состоянию полимера достаточно хорошо аппроксимируются отрезками прямых, а дальнейший нагрев с переходом в вязко-текучее состояние – скорее напоминает криволинейный отрезок. В соответствии с [1], весь характер наблюдаемой зависимости свидетельствует об ионном характере проводимости полимерной матрицы.

## 2. Эксперимент

В наших экспериментах были выполнены исследования температурной зависимости сопротивления сополимера ПВХ-ПАц для различных концентраций двойных полиеновых сопряженных связей (ПСС), определяемых продолжительностью процесса дегидрохлорированния (ПДГХ) методом термолиза ПВХ в растворе [3,4]. Как было показано ранее, в таких сополимерах наблюдаются обратимые, спонтанные и стимулированные внешними воздействиями переключения проводимости из состояния низкой проводимости (СНП) в состояние высокой проводимости (СВП) с перепадом на 11 порядков величины [3,4]. В наших работах [3-6] также наблюдались переходы в образцах ПВХ пленок с использованием пластификатора (модификатора типа А), причем переключения проводимости по порядку величины не превышали 3-4 порядков, но имели также обратимый, спонтанный или стимулированный внешними воздействиями характер.

В данной работе с целью выявления физических механизмов «аномальных» переключений выполнены исследования температурной зависимости в диапазоне от 15 до $85^0C$ электропроводимости для серии образцов сополимеров с различным увеличивающимся содержанием фрагментов двойных связей (от чистого ПВХ до уровня концентраций, при котором получаемый сополимер ПВХ-ПАц допускал формирование пленочных образцов методом полива [1]).

Для количественного подтверждения различной степени ПДГХ исходного ПВХ и корреляции ее со временем термолиза в растворе дополнительно проводилось исследование люминесцентных свойств полученных образцов.

Электрическая схема установки измерения проводимости пленочных полимерных образцов подробно была описана в работах [3,4]. Для измерения температурной зависимости в схему включен модуль изменения температуры образца (термостат) и измеритель собственно температуры рабочей части измерительной ячейки.
В отличие от публикаций [3-6], в данной работе исследовались температурные зависимости для образцов с разным содержанием ПСС, причем контроль относительного содержания двойных сопряженных связей в испытуемых образцах сополимеров проводился методом люминесценции на спектрометре Joben Ivon –U1000, разрешение - $1A^o$, при возбуждении светодиодом с максимумом спектра 365 нм.

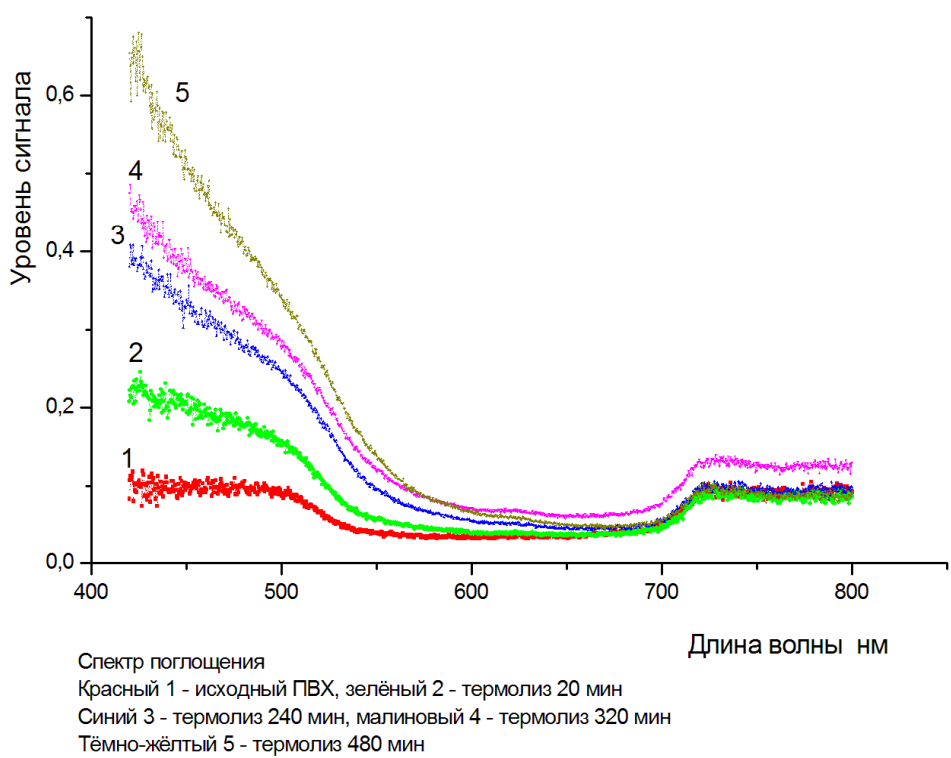

Спектр поглощения
Красный 1 - исходный ПВХ, зелёный 2 - термолиз 20 мин
Синий 3 - термолиз 240 мин, малиновый 4 - термолиз 320 мин
Тёмно-жёлтый 5 - термолиз 480 мин

Рис.1 а. Рост поглощения образцов при увеличении длительности ПДГХ

---

[1] При дегидрохлорировании в растворе максимальная концентрация двойных связей 30%, однако в наших экспериментах, пленочные образцы устойчиво получались при заметно более низких коцентрациях фрагментов молекул полиацетилена.

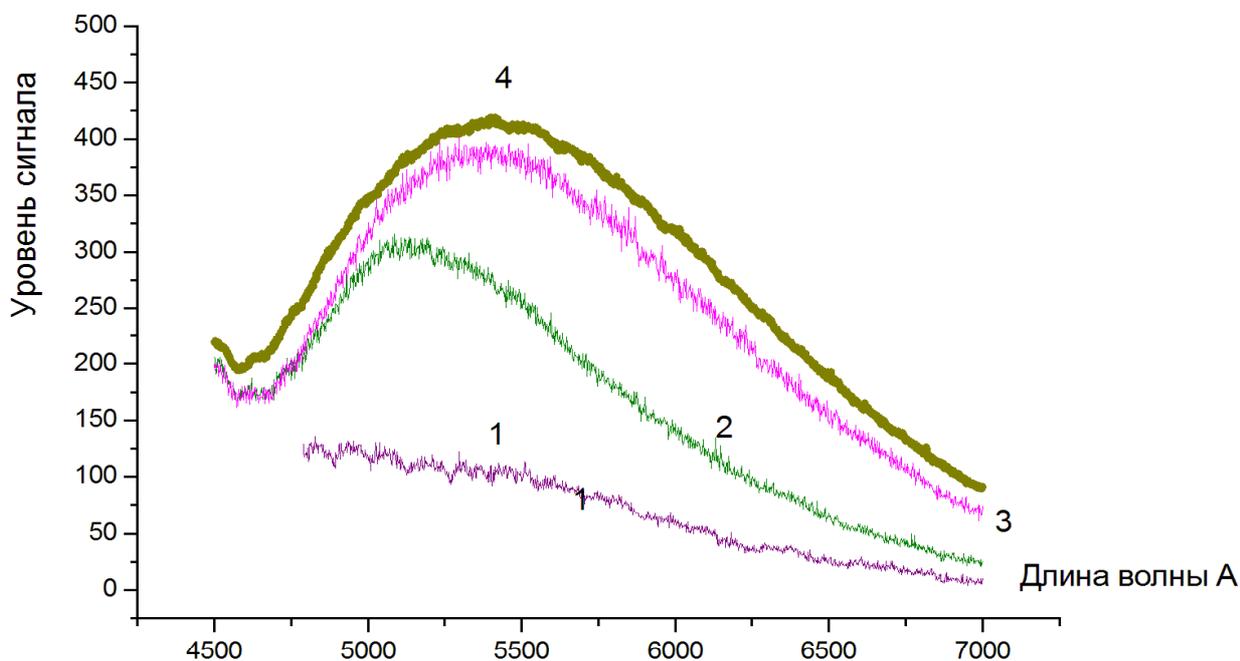

Рис 1 Спектр фотолюминесценции 1-чистый поливинилхлорид (ПВХ)
2-дегидрохлорированный ПВХ термолиз в течение 20 мин
3-дегидрохлорированный ПВХ термолиз в течение 240 мин
4-дегидрохлорированный ПВХ термолиз в течение 320 мин
Термолиз 190 С Длина волны возбуждения светодиодом 365 нм шаг-2 А

На Рис.1 приведены спектры люминесценции исследуемых образцов сополимеров, показывающих, что интенсивность люминесценции пропорциональная количеству двойных сопряженных связей и коррелирует со временем процесса ПДГХ ПВХ для испытуемых образцов. На рис.1 а, спектры поглощения, образцов из ПВХ подвергнутых ПДГХ так же указывают на почти линейный рост поглощения, (т.е. концентрации двойных связей) с длительностью процесса ПДГХ.

### 3. Результаты эксперимента

В наших экспериментах зависимость проводимости образцов сополимеров, как показали эксперименты, не соответствует ни одному из известных перечисленных в кратком введении типов температурной зависимости материалов. Единственная «нормальная» [1] зависимость сопротивления показанная на Рис.2 наблюдается для аналогичного пленочного образца чистого

ПВХ

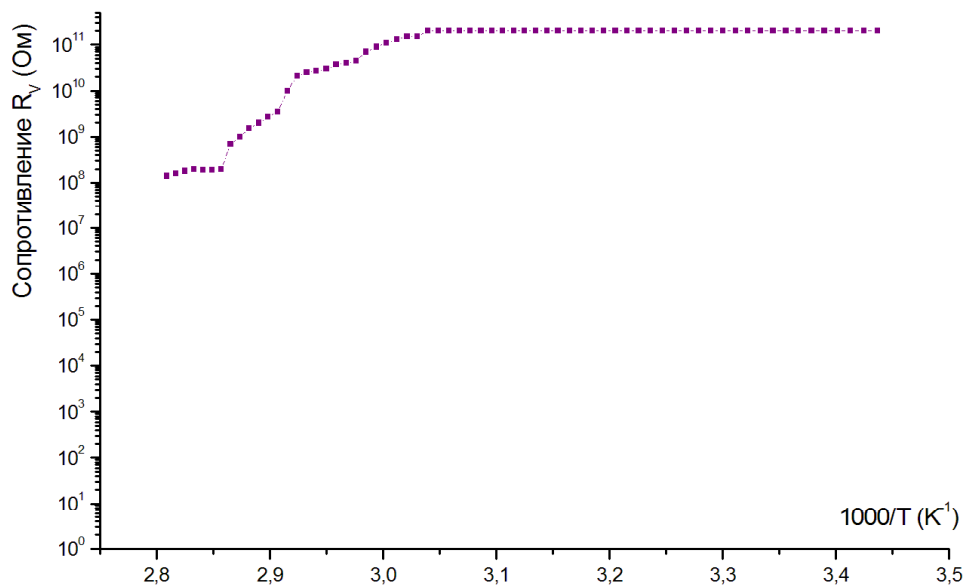

Рис 2 Зависимость объёмного сопротивления $R_V$ (Ом) чистого поливинилхлорида 10 мкм от величины обратной температуры

Ситуация радикально изменяется для даже небольших (композит визуально не отличается от исходного ПВХ) примесей фрагментов молекул полиацетилена. Так для образца с ПДГХ 20 минут (Рис.3) происходит скачок проводимости более чем на 6 порядков величины, с небольшой «ступенькой» в области проводимости порядка $10^6$ Ом. Переключение проводимости происходит «быстрее» экспоненциального закона, и по нашему мнению может быть связано с резким перколяционным переходом из состояния изолятора в полупроводниковое состояние, когда при увеличении температуры энергия термически активированных электронов π-связей становиться достаточной для потенциальных изолирующих барьеров между фрагментами полиацетилена.

Таким образом наиболее вероятным представляется перескоковый механизм между проводящими включениями, причем в нашем случае в качестве проводников могут рассматриваться фрагменты ПАц (содержащего ПСС) в образцах дегидрохлорированного ПВХ.

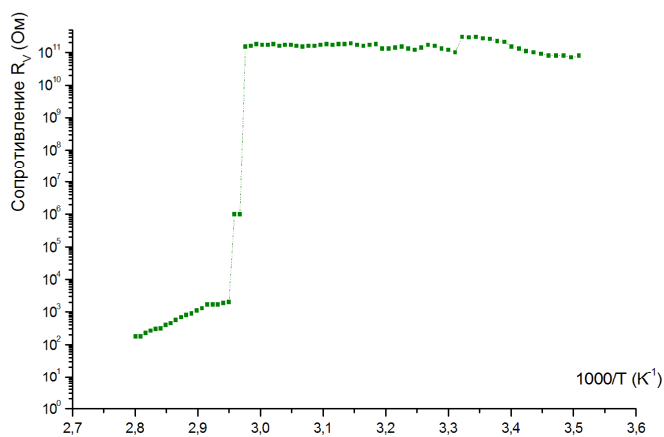

Рис. 3 Температурная зависимость объёмного сопротивления $R_V$ (Ом) дегидрохлорированного поливинилхлорида термолиз 190 С в течение 20 минут

В дальнейшем при изменении температуры образцов (и при приближении к температуре стеклования) происходит еще один скачок, носящий также ступенчатый характер, при котором значения Rv достигают значений единиц Ом. Этот скачок можно связать с возможным переходом через температуру стеклования самого сополимера, когда кинетические сегменты макромолекул, содержащих полярные группы, приобретают значительно большую подвижность (включая дополнительный механизм ионной проводимости [1]). Другой механизм возможного резкого увеличения проводимости может быть связан с локальным саморазогревом токового канала [7] и соответствующим возможным увеличением числа сопряженных связей.

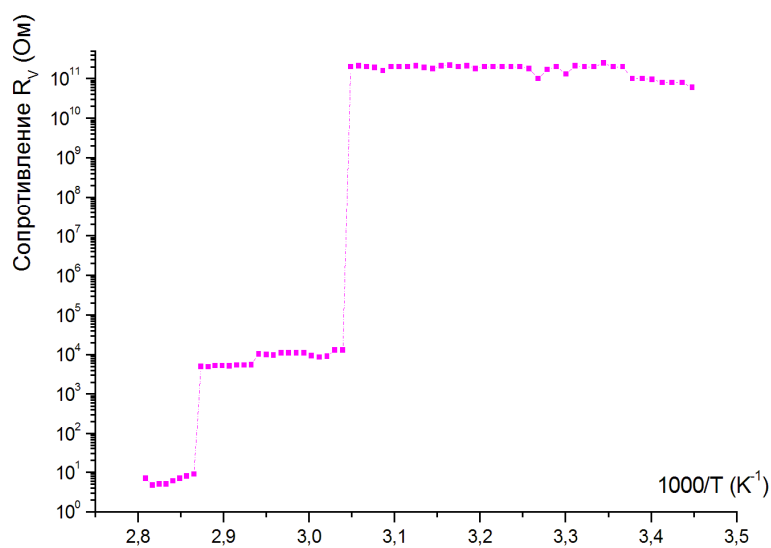

Рис 4 Температурная зависимость объёмного сопротивления $R_V$ (Ом) дегидрохлорированного поливинилхлорида термолиз 190 С в течение 240 минут

Точно такой же двухступенчатый характер изменения электропроводности имеют и образцы с ПДГХ 240 и 320 минут, соответственно (Рис.4,5). Однако здесь значения их $R_v$ на первой («полупроводниковой») ступени на 2 порядка меньше, а плато самих ступеней носит более выраженный протяженный характер.

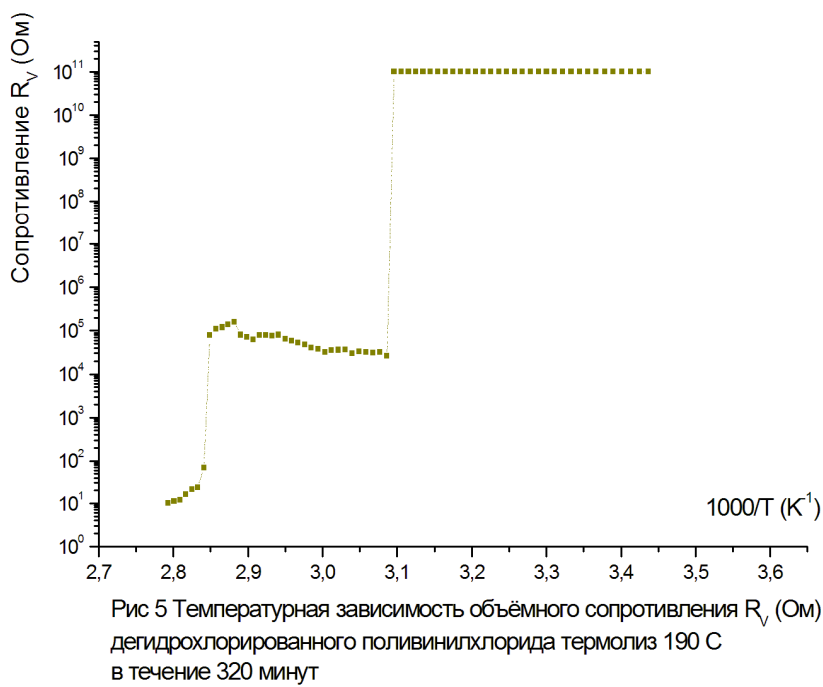

Рис 5 Температурная зависимость объёмного сопротивления $R_V$ (Ом) дегидрохлорированного поливинилхлорида термолиз 190 С в течение 320 минут

При этом первый скачок наступает при более низких температурах. Следует отметить также, что первый («полупроводниковый») скачок происходит при этом при более низких температурах для образцов сополимеров с большим временем дегидрохлорирования, т.е. при большем содержании в них ПСС (см.Рис.6), что на качественном уровне соответствует предложенной модели перколяции прыжковой проводимости.

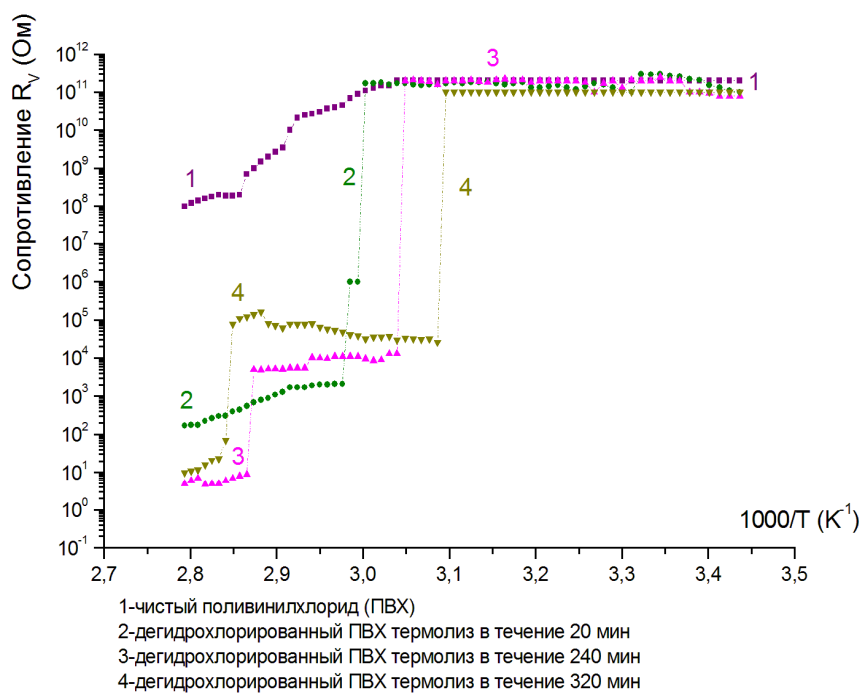

1-чистый поливинилхлорид (ПВХ)
2-дегидрохлорированный ПВХ термолиз в течение 20 мин
3-дегидрохлорированный ПВХ термолиз в течение 240 мин
4-дегидрохлорированный ПВХ термолиз в течение 320 мин

Рис.6. Температурные зависимости сопротивления образцов с разными степенями дегидрохлорирования

## 3. Выводы

Из анализа полученных результатов следует, что:

1. Увеличение доли ПСС (фрагментов ПАц) в макромолекулах исходного ПВХ, образующихся в результате увеличения времени термолиза последнего, хорошо коррелирует с данными по  сперктам поглощения и люминесценции этих же образцов.

2. Образец исходного (чистого) ПВХ при нагреве его в интервале температур стеклования и размягчения (от 15 $^o$ до 85$^o$C) хотя и увеличивает несколько свои электропроводящие свойства (на 2-3 порядка), остается однако в классе изоляторов.

3. Образцы же ПВХ, подвергнутые частичному дегидрохлорированию (т.е. содержащие ПСС) способны переходить при увеличении температуры из СНП в СВП.

4. Переход образцов сополимеров ПВХ-ПАц из СНП в СВП при увеличении температуры носит двухступенчатый характер и проявляется более ярко при увеличении концентрации фрагментов ПСС в образцах сополимеров ПВХ-ПАц. При этом переход носит перколяционный (скачкообразный) характер.

5. На первой ступени перехода (со значениями Rv  ~$10^4$-$10^5$Ом) образцы сополимеров переходят из класса изоляторов условно в класс «полупроводнков».  При этом, чем большее количество ПСС содержит образец, тем при более низкой температуре происходит этот переход..

6. Вторая ступень перехода тоже носит скачкообразный характер, но уже в области температур, близких к температуре стеклования. При этом значения Rv  достигают уже единиц Ом. .

## Литература

1. Б.И. Сажин Электрические свойства полимеров. Л.:Химия, 1977.

2. Э.Р Блайт., Д. Блур.  Электрические свойства полимеров..М.:Физматлит, 2008.

3. Д.В.Власов, Л.А.Апресян, Т.В.Власова, В.И.Крыштоб.
   Аномалии  и пределы точности измерений электропроводности  в
   пластифицированных прозрачных поливинилхлоридных пленках //
   Высокомолекулярные соединения, Серия А, 2011, том 53, № 5, с.739.

4. V. Vlasov, L. A. Apresyan, T. V. Vlasova and V. I. Kryshtob
   Recent plasticizers and conductivity anomalies in homogeneous antistatic transparent
    plasticized PVC films. Ch.6  in:Polymer Relaxation, eds. P.J.Graham and C.M.Neely, Nova
    Science Publ.,2011.

5. Д.В.Власов, Л.А.Апресян, Т.В.Власова, В.И.Крыштоб .Переключения электропроводности пленок пластикатов поливинилхлорида под воздействием одноосного давления. // ЖТФ, 2011, том 81, выпуск 11,с.94-99.

6. D. V. Vlasov, L. A. Apresyan, T. V. Vlasova, V. I. Kryshtob, 2011; On Anomalies of Electrical Conductivity in Antistatic Plasticized Poly(vinyl chloride) Films, American Journal of Materials Science, vol. 1(2): 128-132.

7. D. V. Vlasov, V. I. Kryshtob,L. A. Apresyan, T. V. Vlasova, S. I.Rasmagin.

   An Approach to the interpretation of spontaneous transitions into a state of high conductivity in the PVC composite films// arXiv.org 1312.3435.